\documentclass[11pt]{article}

\usepackage{amsmath,amssymb,amsthm}
\usepackage{geometry}
\usepackage{hyperref}
\usepackage{cite}
\usepackage{bbm}
\usepackage{mathtools}

\title{\large\bf The Awada-Gibbons-Shaw Algebra in de Sitter Space and SUSY Breaking}

\author{T.~Banks\\[4pt]
\small NHETC and Department of Physics and Astronomy\\
\small Rutgers University, Piscataway, NJ 08854}

\date{}

\begin{document}
\maketitle

\begin{abstract}
We rederive the Cosmological Supersymmetry Breaking Relation $m_{3/2} = \frac{C}{\sqrt{R_{dS} L_P}}$ from a deformation of the Awada-Gibbons-Shaw local supersymmetry algebra.  
\end{abstract}

\tableofcontents

\section{Introduction}
More than $40$ years of work in perturbative superstring theory and the AdS/CFT correspondence point to the validity of two conjectures that the were made many years ago:
\begin{itemize}
\item Models of Quantum Gravity are holographic\cite{thooft}\cite{susskind}.  Well established models have quantum degrees of freedom that exist only at the asymptotic boundaries of asymptotically flat or anti-de Sitter (AdS) space-times.  A more local version of this conjecture\cite{ted95} is that the degrees of freedom describing a causal diamond live on the maximal area surface on the diamond's boundary, with a modular Hamiltonian whose expectation value is $\frac{A_{\diamond}}{4G_N}$.  
\item In order to have space-time curvature radius $\gg$ microscopic length scales, the model must be exactly supersymmetric\footnote{I first made this conjecture in 1999, based on the fact that all attempts to find non-SUSY string theories in flat spacetime above two dimensions had failed and that conjectures about SUSY violating AdS/CFT models with large radius had obvious loopholes.  Much later, the conjecture was revived by Ooguri and Vafa, and, as far as I know there are no counterexamples. As I pointed out, for AdS/CFT models constructed from D-branes, it's equivalent to a mathematical conjecture about the finiteness of K-theory groups.} .  For asymptotically AdS space this can be relaxed to include models that are relevant perturbations of superconformal field theories.
\end{itemize}

For negative c.c., the boundary theory is a (relevant perturbation of) a superconformal quantum field theory (SCFT).  For vanishing c.c., $d > 4$ and finite energy, the boundary correlators are described by a unitary scattering operator on a Fock space that is the completion of the space of states of a finite number of super particles.  Super-particle states can be described by the following two algebras defined on the null cone, $P^2 = 0$\cite{ags}\cite{susyscattering}.  
\begin{equation}   [Q_a^{\pm\ I} (P), Q_b^{\pm\ J} (P)]_+ = (\gamma^m)_{ab} P_m \delta^{IJ} . \end{equation}
\begin{equation} (P_0 \pm \vec{P}) \cdot (\gamma^0 , \vec{\gamma})_{ab} Q_b^{\pm\ J} (P) = 0 . \end{equation}  For scattering theory there's one copy of this algebra on the past null cone ($P_0 > 0$) and one on the future null cone ($P_0 < 0$).  For massless particles we only have one orientation of $\vec{P}$ for each sign of $P_0$ corresponding to incoming or outgoing null momenta.  For massive particles we have both orientations of generator in both past and future, and the mass appears in the anticommutator of the generators of opposite orientation.  This is familiar from the mass formulae for BPS particles for extended SUSY.  

For $d = 4$ there is no unitary S operator on Fock space, and in any dimension the gravitational S-matrix fails to be unitary in the infinite energy limit\cite{tbsmatrix}. Both high energy gravitational bremstrahlung and black hole formation and evaporation produce states of soft gravitons that become orthogonal to every Fock space state as $E \rightarrow \infty$ for any initial Fock space state with total energy $E$.  

In\cite{susyscattering} we proposed a definition of gravitational scattering in terms of a state on the Awada-Gibbons-Shaw (AGS) asymptotic SUSY algebra described below.  In\cite{tbsmatrix} we argued that in eleven dimensions that state was the direct limit of the analog of a locally normal state\cite{locallynormal} on a sequence of finite dimensional algebras of fermionic oscillators.  That argument probably goes through for any supersymmetric compactification of string theory for $ d > 4 $, assuming that the perturbation series defines a unitary scattering matrix on Fock space at finite energy.  

The AGS algebra is simply
\begin{equation}   [Q_a^{\pm\ I} (P), Q_b^{\pm\ J} (P^{\prime})]_+ = \delta (P\cdot P^{\prime})(\gamma^m)_{ab} P_m \delta^{IJ} . \end{equation}
\begin{equation} (P_0 \pm \vec{P}) \cdot (\gamma^0 , \vec{\gamma})_{ab} Q_b^{\pm\ J} (P) = 0 . \end{equation} 

The state is defined by insisting that the variables $Q_a^{\pm\ I} (P = 0)$ vanish only on a finite set of annuli on $S^{d-2}$.  For non-zero $P$, $Q_a^{\pm\ I} (P)$, is non-vanishing only in the spherical caps inside of the annuli.  The $d - 2$ volume ({\it area} by abuse of language) of the annulus surrounding a cap with a non-zero value of $P$ is proportional to $P_0 / M_P$ .  

Significantly, this algebra describes any state with a finite number of particles in the supergravity (SUGRA) Fock space, purely in terms of fermionic fields on the null cone. Furthermore, the scale on the null cone enters only through the constraints defined by the state.  Most of the operators $Q_a^{\pm\ I} (P)$, with non-zero $P$,  have vanishing correlation functions in this state, and they are effectively defined by constraints on the operators with $P = 0$ (which nonetheless have non-trivial dependence on $S^{d-2}$. ).  In\cite{tbsmatrix} we showed how to obtain this state, including all of the zero momentum operators, as a limit of a state on an algebra of a finite number of fermionic oscillators. 

\section{Supergravity on a Diamond in Minkowski Space}

The AGS algebra discussed in the previous section describes the entire asymptotic content of supergravity theories in perturbative Fock space for $d > 4$ and finite energy, and the description of the special state on the algebra is a conjecture about how to complete that scattering theory to take into account soft particle states that are encountered in $d =4$ or the limit of infinite energy, which do not correspond to normalizable vectors in Fock space.  The algebra can be viewed as the asymptotic algebra of SUSY currents.  Conventionally\cite{ags} this closes on the Bondi-Metzner-Sachs\cite{BMS} algebra of commuting local momenta, but by realizing the eigenvalues of the BMS algebra as the space on which the SUSY currents live, we trivialize that part of the algebraic structure.  The same algebra of currents is realized by a free gravitino field. In this section, we exploit that fact to explore what happens to the algebra when we restrict attention to a finite causal diamond.  By necessity, this requires a conjecture, and we are guided by two principles.  The first is the conjecture of Carlip and Solodukhin\cite{carlip}\cite{solo} that the effective theory on the boundary of a causal diamond is a $1 + 1$ dimensional conformal field theory (CFT) with central charge proportional to $\frac{A_{\diamond}}{4G_N}$.  This conjecture has received an enormous amount of support\cite{deboeretal}\cite{VZ2}\cite{BZ}\cite{tbpddS}. In particular, in\cite{BZ} it was shown that it was compatible with a cutoff on the CFT, as long as the cutoff was above the steepest descent point that dominates the integral over the Cardy spectrum for large central charge.

The second conjecture\cite{tbwfhst}\cite{tbjk} is that the CFT consists of 1+1 dimensional free fermions\footnote{The fast scrambling properties of horizons and the expected spectral properties of the modular Hamiltonian of black holes require one to add a marginal current current perturbation to the CFT, but that is irrelevant for the purposes of the present paper.}, in one to one correspondence with a cutoff set of spinor spherical harmonics on the two sphere. We will see that precisely this structure arises from the free gravitino field with an angular momentum cutoff, in the vicinity of the maximal area surface on the diamond.  It's of course obvious that if we make an angular momentum expansion of the Rarita Schwinger field, then for fixed values of the angular momentum quantum numbers, we get a two dimensional spinor field propagating in the null coordinates $x^{\pm} = (r \pm t)/2 $.  If we impose boundary conditions, which do not allow energy and momentum to leak out of the diamond, and concentrate on the region near the maximal area surface on the diamond, then we find that the terms involving one of the null derivatives $\partial_{\pm}$ dominate the equation and the equation is approximately the massless $1 + 1$ dimensional Weyl equation.

The SUSY transformations for free gravitions are just inhomogeneous fermionic shifts $\psi_m \rightarrow \psi_m + \partial_m \epsilon,$ so the AGS generators are just components of the gravitino fields themselves.  Smearing the generators with test 
functions $f^{\pm} (x^{\pm})  $ that both have overlap with the maximal area point $x^{\pm} = 0$, we find
\begin{equation} [Q_a^+ (\Omega_1 ) , Q_b^- (\Omega_2) ]_+ = \delta^2 (\Omega_1 - \Omega_2) [(\gamma \cdot n_-)_{ab} P_+ (\Omega_1) + (\gamma \cdot n_+)_{ab} P_- (\Omega_2)] . \end{equation} $n_{\pm}$ are the two null directions, and $P_{\pm} (\Omega)$ are the densities of null momenta.  We will restrict attention to $ d = 4$, as we've done in the above equations, since this is the only dimension where supergravity theories can give rise to de Sitter solutions.   Since the RHS pairs left helicity gravitinos with right moving momenta, and vice-versa, it is an effective mass term in the sense that it transforms as a scalar under Lorentz boosts along either past or future diamond boundaries.  It's of course well known that there are no boundary conditions on finite volume spaces with boundary, that preserve chiral symmetries.

If we now turn our attention to the AGS algebra in de Sitter space we can note several important distinctions. Most importantly, we do not have to impose artificial boundary conditions to prevent information flow out of the maximal dS causal diamond.  Causality prevents it.  Correspondingly, the ingoing and outgoing null momenta near the bifurcation surface, indeed near the entire cosmological horizon, are just limits of the generator of the same flow: the static Hamiltonian.  They simply have opposite sign.  Thus the anti-commutator between future and past AGS generators on the bifurcation surface becomes
\begin{equation} [Q_a^+ (\Omega_1 ) , Q_b^- (\Omega_2) ]_+ = \delta^2 (\Omega_1 - \Omega_2) [\Gamma_{ab} H] . \end{equation} Here $\Gamma$ is a Dirac matrix with equal numbers of $\pm 1$ eigenvalues. 

We now want to impose the Covariant Entropy Principle, which in this context means we should replace the AGS algebra with some finite regularized version, built out of degrees of freedom that implement the rules for the modular Hamiltonian
\begin{equation} \langle K \rangle = \frac{A}{4G_N} . \end{equation}
\begin{equation} \langle (K - \langle K \rangle)^2 \rangle = \langle K \rangle . \end{equation}  The first of these is simply the Gibbons-Hawking\cite{gh} law, while the second can be motivated by\cite{carlip}\cite{solo}\cite{deboeretal}\cite{VZ2}\cite{BZ}\cite{tbpddS}.  The easiest implementation of these rules is to take the AGS generators to have cut off angular momentum on the two sphere, with a cutoff $L = R_{dS}/L_P$ and to let them be the zero modes of $1 + 1$ dimensional massless Dirac fields (restricted to have $10-20$ momentum modes on an interval), labelled by the same angular momenta. 

The angular momentum cutoff is a ${\it fuzzy}$ cut-off of the space of differential forms on the two sphere, but since $R/L_P \sim 10^{61}$ in the real world, we can, to the accuracy required for this paper, assume that the cut-off breaks the sphere up into pixels of area roughly $L_P^2$ . 

In the de Sitter case, it is even more clear that the right hand side of the anti-commutator of the AGS generators on the past and future horizons acts as some kind of mass term for the gravitino.  Consider a self consistent calculation of the mass of a gravitino bound to a detector at the origin of the static patch.  Such a detector can remain near the origin and function as a good ``von Neumann" detector for times of order $R {\rm ln} (R/L_P)$.  
The gravitino wave function will have a transverse spread in static coordinates of order $ 1/m_{3/2}$ and this is unchanged as one extrapolates from the vicinity of the origin to the horizon. We postulate that the total transverse area over which the wave function is spread is unchanged as one extrapolates from the vicinity of the origin to the horizon.   Note that the gravitino wave function is, from the bulk point of view, spherically symmetric.  The density matrix of the dS vacuum state is invariant under a much larger group of fuzzy area preserving diffeomorphisms.  Thus the quantum information about the gravitino is spread randomly over the huge area of the cosmological horizon and the gravitino mass should be obtained by averaging the right hand side of the AGS anti-commutator over an area of order $m_{3/2}^{-2}$ and then taking into account the redshift of energies from the stretched horizon to the origin.  

Near the horizon, everything is set by the Planck scale, so each pixel has a fluctuating $\pm M_P$ contribution to the anti-commutator, if we assume the quantum state on the horizon has an equal probability for the two eigenvalues of $\Gamma$.  This would follow from time reversal invariance if our universe were actually de Sitter space.  It isn't, but the entropy deficit of our cosmological initial conditions is quite negligible compared to the total entropy of the universe, and there is no apparent reason for those initial conditions to favor one sign of $\Gamma$, so this is a very reasonable guess.  In this case the gravitino mass comes from the fluctuations
\begin{equation} m_{3/2} = C (R/L_P)^{-1} M_P \sqrt{A/L_P^2} = C (R m_{3/2})^{-1} M_P . \end{equation}  The constant $C$ reflects the fact that all of our estimates are just ``order of magnitude".  

This is the same relation that was derived, from quite different considerations in\cite{cosmosusy}.  The current derivation of it is a lot cleaner.  It involves a plausible deformation of the AGS algebra of asymptotically flat space using the no longer controversial assumption that asymptotically de Sitter spaces are describable by finite dimensional operator algebras\cite{tbwfds}\footnote{The assertion of\cite{clpw} that de Sitter space should be described by the hyperfinite Type $II_1$ von Neumann factor has no observable consequences, even if one disregards the {\it a priori} constraints on quantum measurements in dS space discussed in\cite{ckn}\cite{tbwfsp}\cite{tbwhisds}. That algebra can always be approximated with arbitrary precision by a finite set of fermion oscillators, so no experiment could ever distinguish it from a finite dimensional algebra. In addition the conjecture of\cite{clpw} raises the question of what the commutant of that algebra in a representation of it in Hilbert space might mean.}.  Calculations are done on the stretched horizon and we use only obvious geometric properties of gravitino states bound to the origin.  

\section{The Constant $C$ and Relations to A Previous Argument}

We've been deliberately cavalier about the ``order one" factors that are represented by the constant $C$ because of the many unknowns that will go into determining them in the real world.  There are at least three sources of ambiguity that we're aware of.  
\begin{itemize}
\item The most precise way one can imagine to specify a theoretical model of the asymptotically de Sitter universe that we appear to inhabit is to embed it in a sequence of models that converge to a mathematically precise model of quantum gravity in asymptotically flat space.  The evidence from perturbative string theory suggests that there may be many such models.  We can pin down the correct one by matching the experimentally observed particle physics to the outlines of it predicted by that supersymmetric theory.  We may expect the gauge groups, light particle representations, and some features of the superpotentials of the chiral multiplets to be well approximated by the supersymmetric limiting theory.  Phenomenological analysis of the constraints on effective field theories that implement the idea that SUSY breaking is mediated by the horizon\cite{cosmosusy} suggest that we have not yet found all of the relevant low energy fields.   This means that we don't actually know the correct algebraic form of the AGS superalgebra that we're trying to use to calculate the gravitino mass.

\item We must impose a fuzzy cutoff on the super-algebra.  In this paper we've implemented that cutoff in a very crude way, which enabled us to get the scaling law for the gravitino mass.  To calculate the constant $C$ we will have to be much more precise. In addition, the fast scrambling properties of horizons and the expected spectral properties of the modular Hamiltonians of chaotic systems suggest that the modular Hamiltonian of de Sitter space is not exactly that of free $1 + 1$ dimensional fermions.  This might also affect the value of $C$.  

\item The AGS algebra we have written down on the horizon of de Sitter space is just the zero momentum mode of a set of $1 + 1$ dimensional fields on the stretched horizon.  The Covariant Entropy Principle tells us to keep only a finite number of these modes but the precise details of how many depend on Planck scale physics to which we may never have access.  It's not clear how this uncertainty affects the value of $C$.  

\end{itemize} 

The other puzzle we should address is the relation between the present calculation and the hand waving argument for the gravitino mass relation presented in\cite{cosmosusy}.  That argument was based on effective bulk field theory and was supposed to imitate the Nambu-Jona-Lasinio self consistent calculation of chiral symmetry breaking.  One imagined calculating the gravitino mass as a self energy insertion to the gravitino propagator, in which a gravitino line went off to the horizon, interacted with the degrees of freedom there and was reflected back.  Assuming the gravitino mass went to zero with the de Sitter radius, the size of this self energy correction, to exponential accuracy is
\begin{equation}  m_{3/2} \sim e^{ - 2 m_{3/2} R } e^{K A/L_P^2} , \end{equation} where $A$ is the area of the horizon that the gravitino can interact with.  Here we've used the Gibbons-Hawking entropy to estimate the density of states on the horizon.  Obviously $A$ increases as the gravitino mass gets smaller, so both terms behave exponentially like powers of $R$.  In order for the gravitino mass to go to zero we must have
\begin{equation}  A/L_P^2  \leq 2/K  R m_{3/2} .  \end{equation}  If the inequality is strict then we would predict $m_{3/2}$ going to zero exponentially like a power of $R$, but this contradicts the inequality because $A$ certainly blows up when $m_{3/2}$ goes to zero.  So we must have
\begin{equation} m_{3/2} A^{-1} = K/2 (L_P^2 / R) .  \end{equation}  We then argued, using effective field theory that particle propagators did random walks on the horizon, which we took to have Planck scale steps in Planck time.  This leads to $A = L_P / m_{3/2}$ up to another order one constant, since a massive particle can only stay on the horizon for a proper time of order its Compton wavelength, and the same scaling relation.  The real puzzle is the discrepancy between the area taken up by the wave function of the massive particle in our present calculation, and this random walk estimate in effective field theory.  

We do not have a derivation of effective field theory from the holographic formalism presented in this paper.  That formalism produces effective Feynman-like diagrams for scattering amplitudes, but they are more like the diagrams for the gravitational Wilson lines that make up the core of jets in attempts to resolve infrared problems by defining more robust observables than individual particle scattering amplitudes.  In previous work on Cosmological SUSY Breaking, we used the equations of effective SUGRA, $m_{3/2} \sim \sqrt{8\pi} F/M_P$, to go from an expression for the gravitino mass to predictions for the scale of SUSY breaking among non-gravitational supermultiplets.  Here, a fairly large value of $C$ was required in order to respect the experimental bounds.  From the point of view of the present paper, it's important to understand that effective SUGRA relation in terms of algebras on the boundary of the static patch, in order to get a handle on a possible fundamental calculation of $C$.

\section*{Acknowledgments}

The author thanks Gemini for useful discussions of the material in this paper.  

\end{document}